\RequirePackage{fix-cm}
\documentclass[smallextended]{svjour3}       
\usepackage{graphicx}
\usepackage{amsmath,amsfonts,amssymb}
\usepackage[colorlinks=true,linkcolor=black,citecolor=blue,urlcolor=blue]{hyperref}
\usepackage{doi}
\newcommand{\eq}[1]{\begin{gather} #1 \end{gather}}

\newcommand{\vc}[1]{{ \boldsymbol #1 }}

\newcommand{\fr}[1]{\ref{fig:#1}}
\newcommand{\er}[1]{(\ref{eq:#1})}

\newcommand{\sr}[1]{\ref{sec:#1}}

\journalname{Exp. Fluids}

\begin{document}

\title{Real-time quantitative Schlieren imaging by fast Fourier demodulation of a checkered backdrop}

\titlerunning{Synthetic Schlieren by FFT demodulation}        

\author{Sander Wildeman}


\institute{S. Wildeman \at
             Institut Langevin - Ondes et Images, ESPCI, PSL Research University, 1 rue Jussieu, 75238, Paris, Cedex 05, France \\
              \email{swildeman@gmail.com}           
}

\date{Received: \today / Accepted: date}

\maketitle

\begin{abstract}
A quantitative synthetic Schlieren imaging (SSI) method based on fast Fourier demodulation is presented. Instead of a random dot pattern (as usually employed in SSI), a 2D periodic pattern (such as a checkerboard) is used as a backdrop to the refractive object of interest. The range of validity and accuracy of this ``Fast Checkerboard Demodulation'' (FCD) method are assessed using both synthetic data and experimental recordings of patterns optically distorted by small waves on a water surface. It is found that the FCD method is at least as accurate as sophisticated, multi-stage, digital image correlation (DIC) or optical flow (OF) techniques used with random dot patterns, and it is significantly faster. Efficient, fully vectorized, implementations of both the FCD and DIC/OF schemes developed for this study are made available as open source Matlab scripts.

\end{abstract}

\section{Introduction and motivation}
\label{intro}

Over the last few decades, with the advent of high-resolution digital cameras and cheaply available computing power, synthetic Schlieren imaging (SSI) \cite{Sutherland1999,Dalziel2000}, also known as background-oriented Schlieren (BOS) \cite{Meier2002}, has proven to be a versatile tool for the full-field measurement of optical inhomogeneities (Schlieren) in transparent media \cite{Murase1990,Richard2001,Moisy2009,Raffel2015,Letelier2016,Hayasaka2016}. The optical setup, proposed well before the digital era \cite{Schardin1942,Burton1949}, is extremely simple and low-cost, requiring only a digital camera and a suitable backdrop (see Fig. \fr{setup}(a)). Transverse gradients in optical path length between background pattern and camera cause distortions in the pattern as seen by the camera (Fig. \fr{setup}(b-d)). The actual Schlieren image is finally constructed ``synthetically'' by digitally comparing the distorted image to an undistorted reference. For quantitative measurements, SSI boils down to the mathematical inverse problem of finding the vector displacement field $\vc u(\vc r)$ that warps the reference image $I_0(\vc r)$ into the observed distorted image $I(\vc r)$,
\eq{
	I(\vc r) = I_0(\vc r - \vc u(\vc r)), \label{eq:warp}
}
where $\vc r$ denotes the pixel coordinates $(x,y)$ in the images. To first order, $\vc u(\vc r)$ is in turn directly proportional to local gradients in optical path length in the Schlieren object \cite{Schardin1942}.

\begin{figure}
\centering
\includegraphics[width=.75\textwidth]{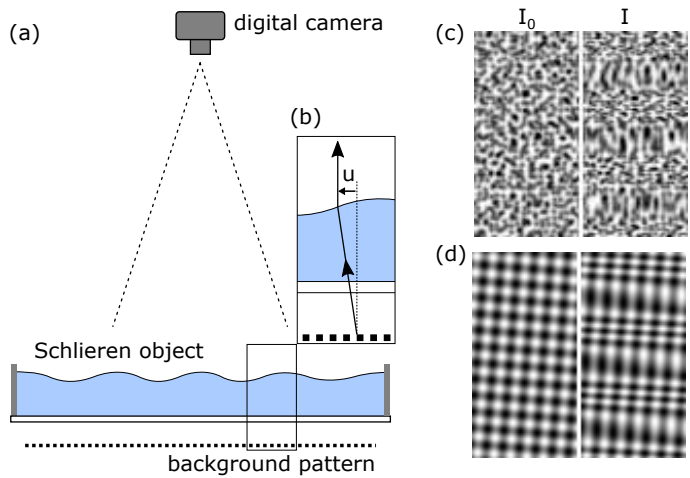}
\caption{Setup for synthetic Schlieren imaging. (a) A digital camera views a textured background through the Schlieren object (here a transparent container filled with water). (b) Optical inhomogeneities in the Schlieren object (e.g., a wave on the water surface) cause light rays between the background and camera to bend, causing the camera to see a distorted pattern. The texture in the background can be taken to be random (c) or periodic (d). In (c) and (d) the undistorted pattern $I_0$ is shown next to the distorted version $I$, for the the same wave-like disturbance.}
\label{fig:setup}
\end{figure}

To attack the Schlieren reconstruction problem, usually computer vision techniques, such as digital image correlation (DIC) or optical flow (OF) are employed \cite{Atcheson2009}. In DIC the reference image is first sliced up into small sub-images, after which, for each sub-image, a match is found in the distorted image by means of cross correlation. The translation vectors obtained in this way constitute a measure of $\vc u(\vc r)$. In OF, on the other hand, one starts from a linearized version of equation~\er{warp}, $I - I_0 = -\vc u \cdot \nabla I_0$, which is then solved for each pixel in the image. Since this system consists of two unknowns $\vc u = (u,v)$ for each one pixel, some form of regularization (or coarse-graining) is required to close the problem (and to make the method robust against noise) \cite{Horn1981}.  An advantage of these vision techniques is that they can be used with almost any type of background \cite{Hargather2010}. The only strict requirement being that the background has sufficient texture at the smallest physical scale to be resolved. In laboratory environments, a random pattern of dots is a common choice \cite{Atcheson2009}.  

However, a drawback of using DIC or OF for distortion tracking is that neither produces a faithful solution to equation \er{warp} in a single pass. In DIC, for example, it is implicitly assumed that the displacement field is approximately constant at the scale of the sub-images. If this condition is violated, local features in $I$ will be elongated (or compressed) with respect to those in $I_0$ (see figure \fr{setup}(c)), giving rise to a large uncertainty in the position of maximum correlation \cite{Meunier2003,Moisy2009}. Furthermore, for detecting sub-pixel displacements with DIC, some form of interpolation of the image data must be performed. As this is inherently in conflict with the aforementioned requirement of having fine texture in the background, the smallest displacement that can be measured with DIC is of the order of one tenth of a pixel in practice. With the OF method one can in principle obtain a dense (per pixel) solution for the displacement field, at a sub-pixel accuracy. However, here it is often difficult to find a good balance between smoothing and overfitting in adjusting the regularization parameter(s). Moreover, because of the linearization, a basic OF algorithm cannot directly handle large distortions. Although many of the above limitations of DIC and OF can be alleviated by incorporating the methods in some sort of recursive (multi scale) warping scheme (see e.g. \cite{Brox2004,Meinhardt-Llopis2013}), only at the expense of speed and simplicity, and with the risk of accumulating errors. 

Here a different approach to the Schlieren reconstruction problem is therefore taken. Instead of adapting a general purpose feature tracking algorithm, it is proposed to employ a special purpose background pattern that allows for a direct inversion of equation \er{warp}. In particular, the idea is to use a regular, 2D periodic (checkerboard-like) pattern with a high spatial frequency (see Fig. \fr{setup}(d)). In analogy with frequency modulation (FM) radio transmission \cite{Carson1922}, this pattern will act as a high frequency carrier wave that is phase modulated by the Schlieren signal. Under well defined conditions (discussed in section \sr{limits}) basic Fourier demodulation techniques can then be used to extract this signal in a straightforward manner.

The idea of extending FM encoding to the spatial domain is not new. It has, for example, successfully been applied in the fields of light projection profilometry \cite{Takeda1982,Takeda1983,Cobelli2009} and digital image strain analysis \cite{Grediac2016}. It has also been tried for Schlieren imaging \cite{Akatsuka2011,Hatanaka2015}. However, in these studies, a 1D stripe motif was employed as a backdrop, instead of the 2D checkered pattern proposed here. While it is true that a 1D pattern has a somewhat larger ``bandwidth'' than a 2D pattern, a disadvantage (in the case of SSI) is that only one component of the displacement field, namely that perpendicular to the stripes, can be recovered from it. As will be shown in the next section, the use of a 2D periodic background pattern allows for a complete reconstruction of the vector displacement field.

This article is structured as follows. First, in section~\sr{method} the spatial Fourier demodulation technique is reformulated in the general language of crystallography, rendering it applicable to any (2D) periodic pattern. From this the theoretical constraints of the method, hereafter simply referred to as FCD, for Fast Checkerboard Demodulation, are derived in section~\sr{limits}. Finally, in section~\sr{comp}, FCD is put to the test in a SSI setting for measuring small waves on a water surface. The practical performance of FCD, in terms of speed and accuracy, is here directly compared to that of a hybrid DIC+OF scheme.

\section{Fourier demodulation of arbitrary 2D periodic patterns}
\label{sec:method}

In Schlieren imaging the quantity of interest, $\vc u(\vc r)$ in equation \er{warp}, is a vector field, consisting of two independent components for each pixel. For a complete Schlieren reconstruction, the background pattern $I_0(\vc r)$ must somehow reflect this fact. In the case of spatial Fourier demodulation, this means that the pattern should contain at least two linearly independent carrier wave vectors.

In crystallography a general 2D periodic pattern is expressed as a harmonic series of the form:
\eq{
	I_0(\vc{r}) = \sum_{m=-\infty}^{\infty} \sum_{n=-\infty}^{\infty} a_{mn} e^{i (m \vc k_1 + n \vc k_2) \cdot \vc r}, \label{eq:refim}
}
where $\vc k_1 \times \vc k_2 \neq 0$ and, since $I_0$ is real valued, the expansion coefficients satisfy $a(m,n) = a^*(-m,-n)$. The vectors $\vc k_1$ and $\vc k_2$ are called the reciprocal lattice vectors. They denote the spatial frequencies (in two directions) of the smallest sub-image, or unit cell, that tiles the plane to form the pattern.

By combining equations \er{warp} and \er{refim}, the distorted image becomes:
\eq{
	I(\vc r) = I_0(\vc r - \vc u(\vc r)) = \sum_{m=-\infty}^{\infty} \sum_{n=-\infty}^{\infty} a_{mn} e^{i (m \vc k_1 + n \vc k_2) \cdot (\vc r- \vc u(\vc r))} \label{eq:defim}
}
From this expression it is clear that the role of the displacement field $\vc u(\vc r)$ is to modulate the phase of each harmonic in the series.

\begin{figure}
\centering
\includegraphics[width=\textwidth]{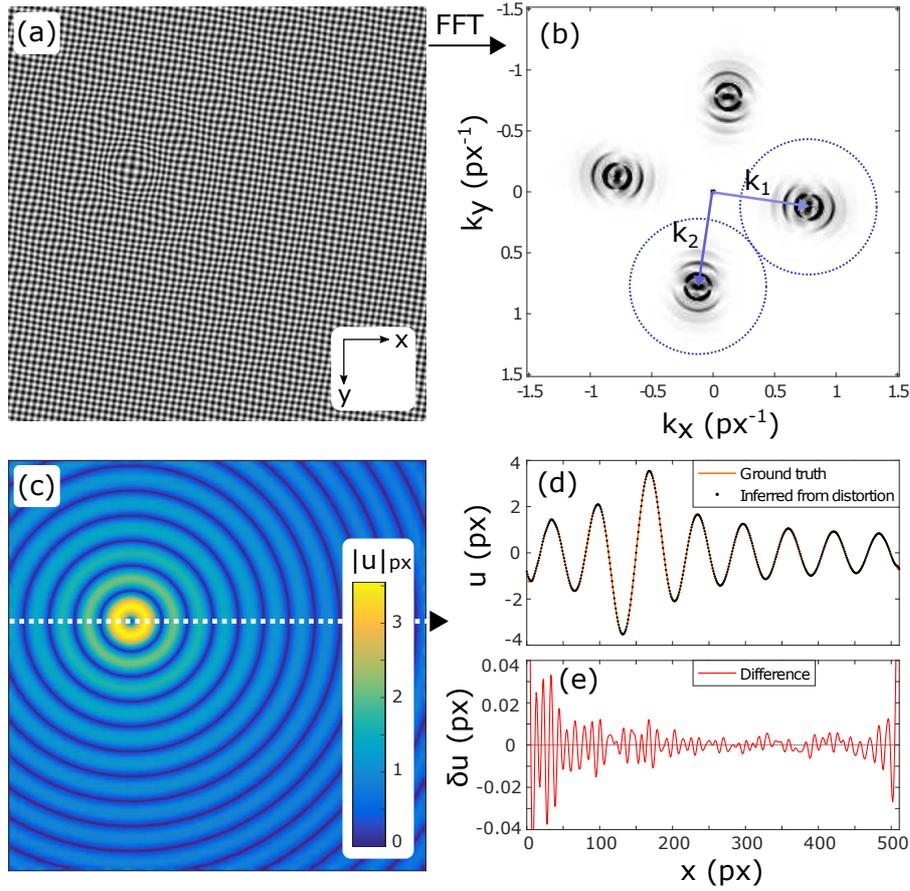}
\caption{Fourier demodulation of a 2D periodic pattern. (a) A checkered background pattern is optically distorted by a small wave like disturbance. (b) In the Fourier domain (k-space) this distorted pattern takes the form of sharp carrier peaks (denoted by $\vc k_1$ and $\vc k_2$) corresponding to the periodic background, which are ``dressed'' (or modulated) by the distortion signal. (c) By extracting the modulated peaks in the Fourier domain (using the circular filter indicated in (b)), the displacement field $\vc u$, representing the distortion, can be reconstructed ($|\vc u|$ is shown here). (d) Displacement in $x$-direction along the dotted line in (c). Dots show the displacement as extracted from (a) and the solid line shows the ground truth. (e) Difference between extracted signal and ground truth.}
\label{fig:fcd}
\end{figure}

Now the basic idea of Fourier demodulation is as follows \cite{Takeda1982,Takeda1983,Cobelli2009,Grediac2016}: if the distortion signal is not too large and the wavelengths comprising it are not too small (as discussed in section \sr{limits}), then the effect of $\vc u(\vc r)$ will be localized around the carrier peaks, $\vc k_c \in n \vc k_1 + m \vc k_2$, in the spatial Fourier domain (k-space). Simple Fourier domain filtering can then be used to single out (from the sums in equations \er{refim} and \er{defim}) complex signals of the form:
\eq{
	g_0(\vc r) = a_c \, e^{i \vc k_c \cdot \vc r},  \\
	g(\vc r) = a_c \, e^{i \vc k_c \cdot (\vc r - \vc u(\vc r))}.
}
Once these signals are extracted, it is a simple manner to obtain $\vc u(r)$ by calculating:
\eq{
	\phi(\vc r) \equiv \text{Im} (\ln (g g_0^*)) = - \vc k_c \cdot \vc u(\vc r). \label{eq:log}
}
By doing this for two linearly independent reciprocal vectors, say $\vc k_1$ and $\vc k_2$, one obtains two equations per pixel,
\eq{
	\phi_1(\vc r)  = -\vc k_1 \cdot \vc u(\vc r) \label{eq:G1}  \\
	\phi_2(\vc r)  = -\vc k_2 \cdot \vc u(\vc r),\label{eq:G2}
}
which can be directly solved for $\vc u (\vc r)$.

In Fig. \fr{fcd} the above procedure is outlined in a graphical way. Figure \fr{fcd}(a) shows a pattern of the form $I_0(\vc r) = \frac 12 + \frac 14(\cos(\vc k_1 \cdot \vc r) + \cos(\vc k_2 \cdot \vc r))$, where in this case $|\vc k_1| = |\vc k_2|$ and $\vc k_1 \perp \vc k_2$. If one looks carefully, one can observe that this checkered pattern is distorted by a small, wave-like, disturbance. In the Fourier domain (Fig. \fr{fcd}(b)) this same image consists of four carrier peaks, from the background pattern, which are ``dressed'' by the signal from the disturbance. Because the disturbance in this example consists of an axisymmetric wave of a single dominant wavenumber $k_s$, the modulation signal takes the form of rings around the carrier peaks, spaced by $k_s$. Notice that these rings fade out in directions orthogonal to the carrier wave vectors, reflecting the reduced distortion sensitivity of the underlying 1D fringe patterns in those directions. Since the modulated Fourier peaks in Fig. \fr{fcd}(b) have practically no overlap, they can be separated out without ambiguity by means of the indicated circular filters. Finally, Fig. \fr{fcd}(c) shows the displacement field obtained after transforming the filtered signals back to the spatial domain, applying equations \er{G1} and \er{G2} and solving for $\vc u(\vc r)$. As demonstrated in Figs. \fr{fcd}(d)-(e), the match between the actual signal and the reconstructed signal is close to perfect for this synthetic example.

In the above description, some important difficulties were glossed over. For one, the complex logarithm in equation \er{log} is multivalued. In most software implementations it has a branch cut on the negative real axis, so that the imaginary part is constrained to lie in between $-\pi$ and $\pi$. Values outside this range are ``wrapped'' around, leading to jumps of $2\pi$ in the extracted phase maps. Furthermore, one may wonder what are the precise criteria for the modulated carrier peaks to be ``well separated'' in k-space. In the next section, these matters are discussed in some detail, with the purpose of giving practical guidelines for the design of the background pattern.

\section{Pattern design and measurement range}
\label{sec:limits}

In this section several theoretical criteria for faithful Schlieren reconstruction with FCD are derived. The aim is to give practical guidelines for the design of the background pattern, given the frequency content and amplitude of the physical signal, and the resolution of the camera.

\begin{figure}
\centering
\includegraphics[width=\textwidth]{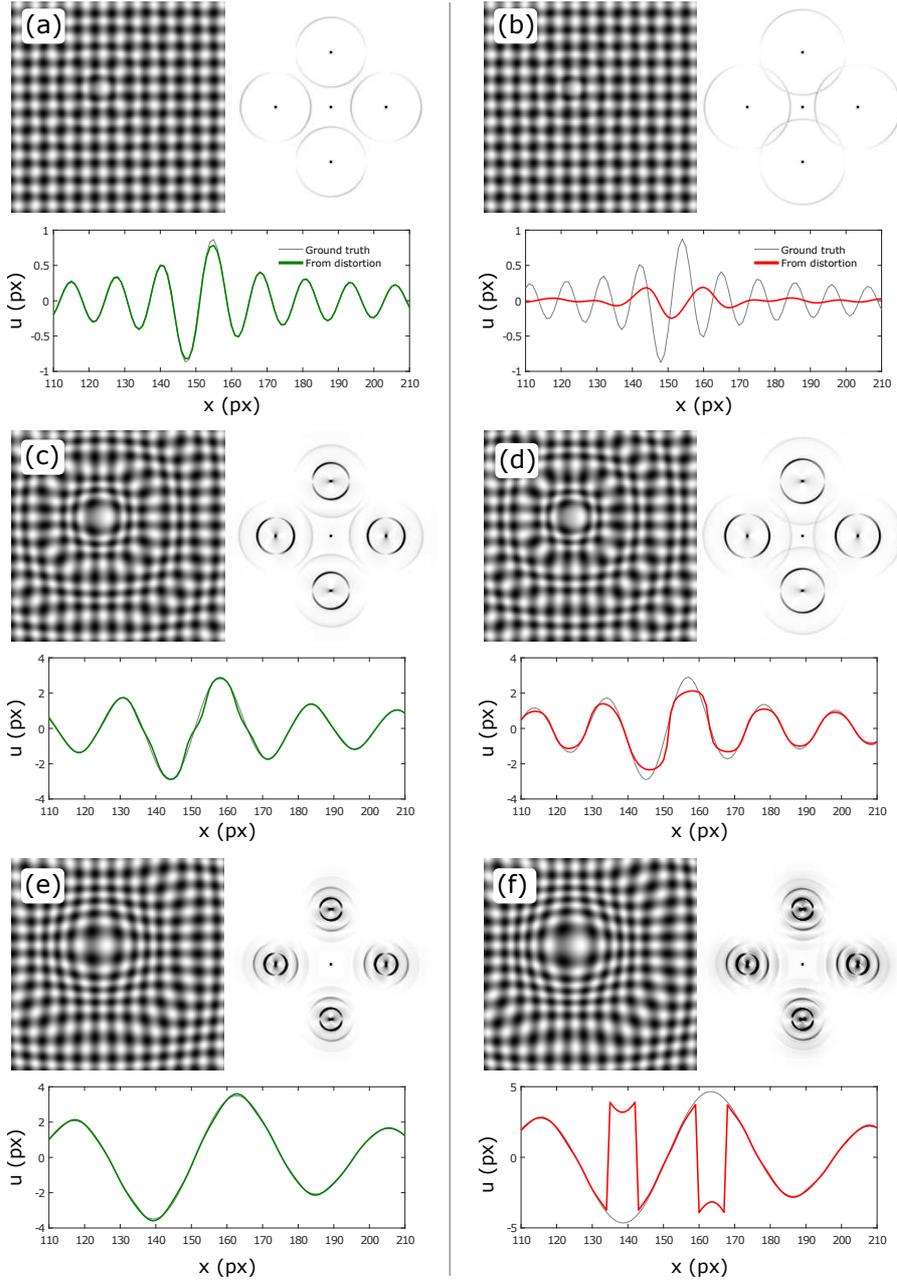}
\caption{Theoretical criteria for faithful reconstruction of the Schlieren displacement field. In the cases on the left (a, c, e) all of the required conditions are (just) satisfied, while in each of the situations on the right (b, d, f), precisely one criterion is just violated. In each case the raw, distorted image, its Fourier transform and the extracted displacement field (along a horizontal trace through the point of maximum distortion) are shown. In (b) the signal wavelength is too small compared to pattern wavelength, $k_s > \Delta k_c/2$. In (d) the modulated carrier signals partially overlap in Fourier space, $k_c k_s u_s > \Delta k_c/2$. In (f) phase wrapping occurs, $k_c u_s > \pi$.}
\label{fig:valid}
\end{figure}

\subsection{Preventing overlap of modulated carrier peaks}

To be able to extract the distortion signal from the carrier background pattern using Fourier domain filtering, the modulated carrier peaks must be well separated in k-space (c.f. Fig. \fr{fcd}(b)). For this to be the case, two conditions have to hold. First, the smallest relevant wavenumber in the physical signal $k_s$ should be smaller than half the distance between the carrier peaks $\Delta k_c$ in k-space, that is:
\eq{
	k_s < \frac 12 \Delta k_c. \label{eq:critks}
}
As illustrated in Fig. \fr{valid}(a-b), wave components in the signal that violate this condition are effectively filtered out in the reconstruction. At the same time signal leaks in from the neighboring peaks, causing aliasing artifacts.

Second, the above condition should also hold for any significant harmonics of $k_s$ in the modulated carrier signal. In the Fourier representation, these harmonics show up as secondary rings around the carrier peaks (see Figs. \fr{valid}(c-d)). The radius (in k-space) of the envelope that encompasses the relevant signal can be estimated by considering the effective (local) wave vector of the modulated carrier signal, $\vc k_{eff} = -i\nabla g/g = \vc k_c + \nabla ( \vc k_c \cdot \vc u )$. For a physical signal component with amplitude $u_s$ and wavenumber $k_s$ this leads to the criterion:
\eq{
	k_c k_s u_s < \frac{1}{2}\Delta k_c, \label{eq:critus}
}
where $k_c \equiv |\vc k_c|$ denotes the magnitude of the carrier wave vector. Basically, equation \er{critks} and \er{critus} constitute the spatial analog of Carson's bandwidth rules for FM transmission \cite{Carson1922}.

In the special case of a square lattice (like a checkerboard) we have $\Delta k_c = \sqrt2 k_c$, so that conditions \er{critks} and \er{critus} can be simplified to:
\eq{
	 \frac{k_s}{k_c} <  \frac{1}{\sqrt{2}}, \label{eq:critkssq} \\
	 k_s u_s < \frac{1}{\sqrt{2}}.
}
Roughly, the first criterion says that the lattice spacing in the background pattern should be smaller than smallest wavelength in the physical signal, while the second condition puts a limit on the magnitude of the virtual strain (gradients in $\vc u$) that can be resolved (independent of the lattice spacing).  Interestingly, this theoretical maximum strain is very close to what can be practically achieved with correlation based techniques \cite{Moisy2009}. It is somewhat more restrictive than the invertibility limit, $k_s u_s < 1$, above which ray crossings may occur \cite{Moisy2009}. In practice it means that the distance between the background pattern and the Schlieren object should be decreased until the sensitivity is small enough for condition \er{critus} to hold (for the smallest wavelengths in the signal).

\subsection{Phase wrapping}

An additional complication in the reconstruction is that the imaginary part of the complex logarithm (Eq. \er{log}) wraps around with a period of $2\pi$, causing the extracted phase to have discontinuities at locations where it exceeds $\pi$  ($3 \pi, 5 \pi$, etc.) in magnitude (see Fig. \fr{valid}(e-f)). For phase wrapping to be absent, the following condition should hold everywhere:
\eq{
	k_c u_s < \pi
}
For a square lattice, this means that the distortions should remain smaller than half the lattice spacing. If the distortions are larger, the phase needs to be unwrapped before solving for $\vc u(\vc r)$. This is not a trivial problem for 2D phase fields with measurement noise. Fortunately, various good algorithms have been developed over the years for this purpose. For example, the 2D-SRNCP unwrap algorithm proposed by Arevalillo-Herr\'aez \textit{et al.} \cite{Herraez2002} and the deterministic unwrap scheme of Volkov and Zhu \cite{Volkov2003} were found to do a good job on the experimental Schlieren data.

In case distortions are tracked over time (using a video camera) another option is to unwrap the phase maps pixel by pixel in the time dimension (or to perform the FCD analysis between subsequent movie frames, so that the incremental displacements stay within the limit).

\subsection{Limits imposed by finite image resolution}

According to condition \er{critks} it is advantageous to design a pattern with a wavelength which is as small as possible, so that the peak spacing in the Fourier domain is as large as possible. However, the finite sampling resolution of the digital image sensor of course puts a lower bound on this. According to the Nyquist-Shannon sampling criterion, aliasing occurs for a perfect sine wave when its period is smaller than two times the sampling rate, that is, smaller than 2 image pixels. To additionally prevent aliasing of the superimposed modulation signal, a minimum pattern wavelength of $2+\sqrt{2}\approx 3.4$ px would be required  for an ideal sine wave background pattern (such as in Fig. \fr{fcd}). However, it is often not easy to produce a pure sine wave background (one needs a properly calibrated printer for example), so that significant signal is present in higher order carrier peaks (i.e., $a_{mn} \neq 0$ for $m,n > 1$ in Eq. \er{refim}). In practice, pattern wavelengths above about 6 px work well. Together with the magnification of the optical setup and the camera pixel size, this sets (through condition \er{critks}) the smallest physical wavelength that can be resolved.

\section{FCD in practice}
\subsection{A one to one comparison to DIC}
\label{sec:comp}

To test the FCD method in practice, and to compare its performance to the more conventional DIC/OF approach, a simple SSI system was set up to measure small waves on a water surface \cite{Murase1990,Moisy2009} (c.f. Fig. \fr{setup}(a)). In this case, the observed image displacements are (to first order) directly proportional to the local slope of the water surface, $\vc u(\vc r) \approx -H (1- n_a/n_w) \nabla h(\vc r)$, where $h(\vc r)$ is the surface elevation, $H$ is the distance between the water surface and the background pattern, and $n_a/n_w \approx 0.75$ is the refractive index contrast between air and water \cite{Murase1990,Moisy2009}.

\subsubsection{Experimental procedure}

\begin{figure}
\centering
\includegraphics[width=\textwidth]{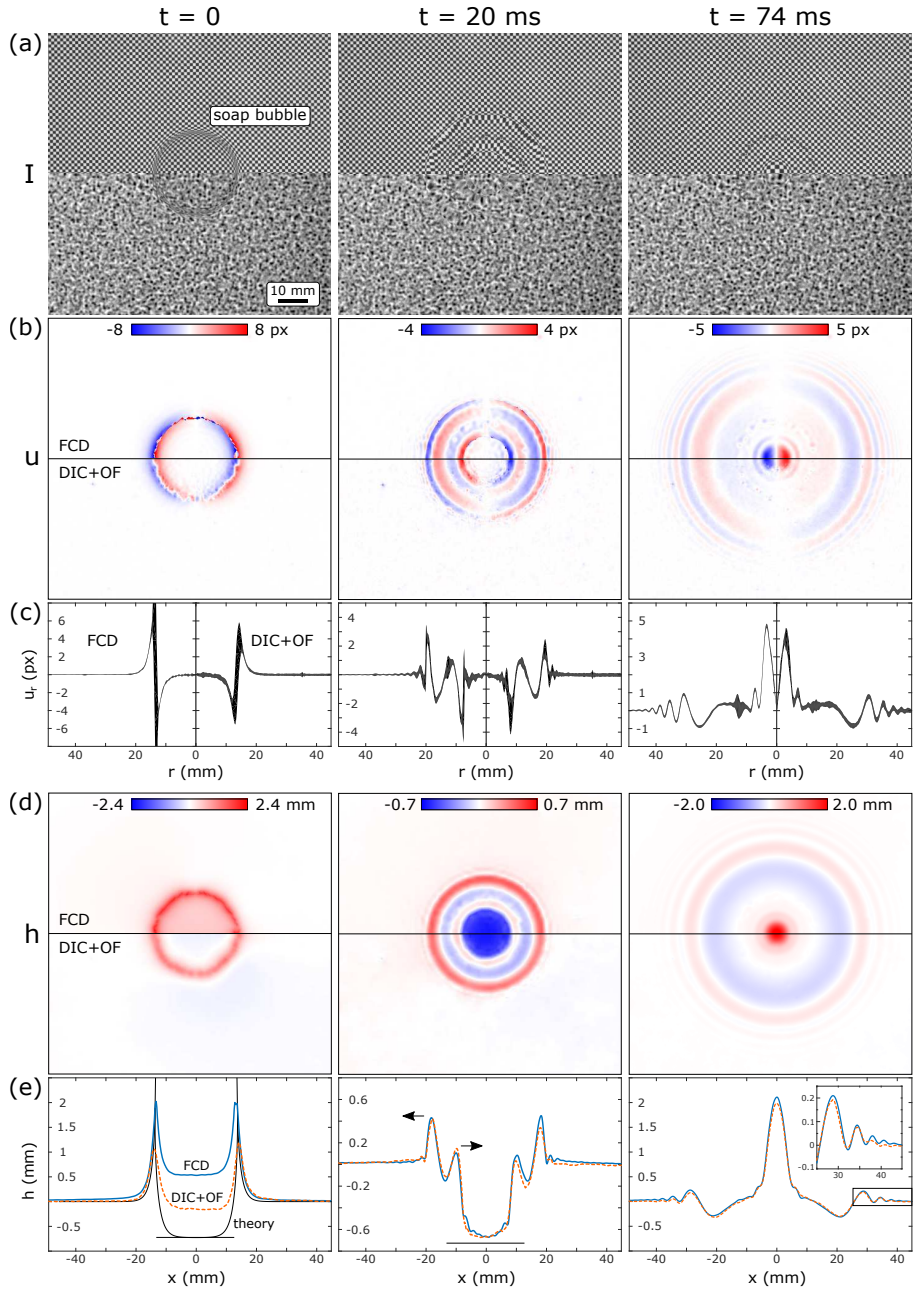}
\caption{Comparison between the FCD and DIC+OF method for synthetic Schlieren imaging of waves on a water surface, generated by a popping soap bubble. (a) Synchronized and aligned video footage of two popping soap bubbles seen from the top. In the first video (top half), a checkerboard pattern was placed on the floor of the container, while in the second video (bottom half) a random dot pattern was used. Snapshots at $t = 0$ (just before the bubble pops), $t = 20\,$ms and $t=74\,$ms are shown. (b) The checkered images were analyzed using the FCD method and the random dot images were processed with a hybrid DIC+OF scheme. Only the $x$ component of the full displacement field is shown here. (c) Radial displacements $u_r$, averaged over 100 equispaced line traces emerging from the center of the bubble. The thickness of the lines is equal to four standard deviations (calculated for each radial position). (d) Surface height fields $h(\vc r)$ obtained by integrating the displacement fields. The correct physical units are obtained by applying a conversion factor derived from Snell's law. (e) Cross-sectional views (through the center of the bubble) of the measured water height.  At $t=0$ ,the theoretical profile is also shown (thin black line). The horizontal lines at $t=0$ and $t = 20\,$ms denote the theoretical indentation due to the Laplace pressure inside the bubble. The arrows at $t = 20\,$ms indicate the direction of propagation of the waves. An inset at $t = 74$ shows a magnified view the small outward propagating capillary waves.}
\label{fig:comp}
\end{figure}

To generate feature rich wave patterns for this test, centimetric hemispherical bubbles were inflated on the water surface. A small amount of dishwashing detergent was added to the water to stabilize the bubbles. Surface tension in the soap film makes that these floating soap bubbles pull the water surface up at their edge, while at the same time pushing it down in the central region (because of the Laplace pressure inside bubble). When the bubble bursts, this static equilibrium shape is suddenly released, setting the initial condition for waves that propagate radially inwards and outwards from the raised rim. For a given bubble size (controlled through the injected volume of air), the waves produced in this way turned out to be very reproducible. 

In Fig. \fr{comp}(a), snapshots of video recordings of two such bubble-popping experiments are shown. In the first experiment, a checkered pattern was taken as a background (for the FCD analysis), while in the second experiment, a random dot pattern was used (for the DIC/OF analysis).  Both events were recorded at a frame rate of 500\,Hz and synchronized so that $t = 0$ corresponds to the moment just before the bubble pops (with an accuracy of $1/500 = 2\,$ms). To facilitate a direct comparison, the image frames were cropped and aligned so that the top half of the first bubble matches up with the bottom half of the second.  The background patterns were printed on fine translucent tracing paper using a laser printer with a resolution of 600\,DPI and illuminated from the back with an uniform white light. In both experiments, the distance between the pattern and the water surface was $H = 8\,$mm. Notice that instead of the ideal interlaced sine wave pattern, a checkerboard pattern was here used with the FCD analysis. The main practical advantage of using a checkerboard pattern is that, due to its binary nature, it is relatively easy to generate at high resolution with any type of (low end) printer. The Fourier transform of a checkerboard pattern has peaks only at locations $\vc k_{mn} = m \vc k_1 + n \vc k_2$, where $m+n$ is odd. This, together with the fact the strength of these peaks decreases roughly as $1/|\vc k_{mn}|$, limits the aliasing artifacts discussed in the previous section. The strength of the secondary peaks can be further reduced by slightly defocussing the pattern, which effectively acts as an optical low-pass filter \cite{Su1992,Lohry2012,Fu2013}.

\subsubsection{Implementation of the distortion tracking algorithms}
\label{sec:impl}

\begin{figure}
\centering
\includegraphics[width=.8\textwidth]{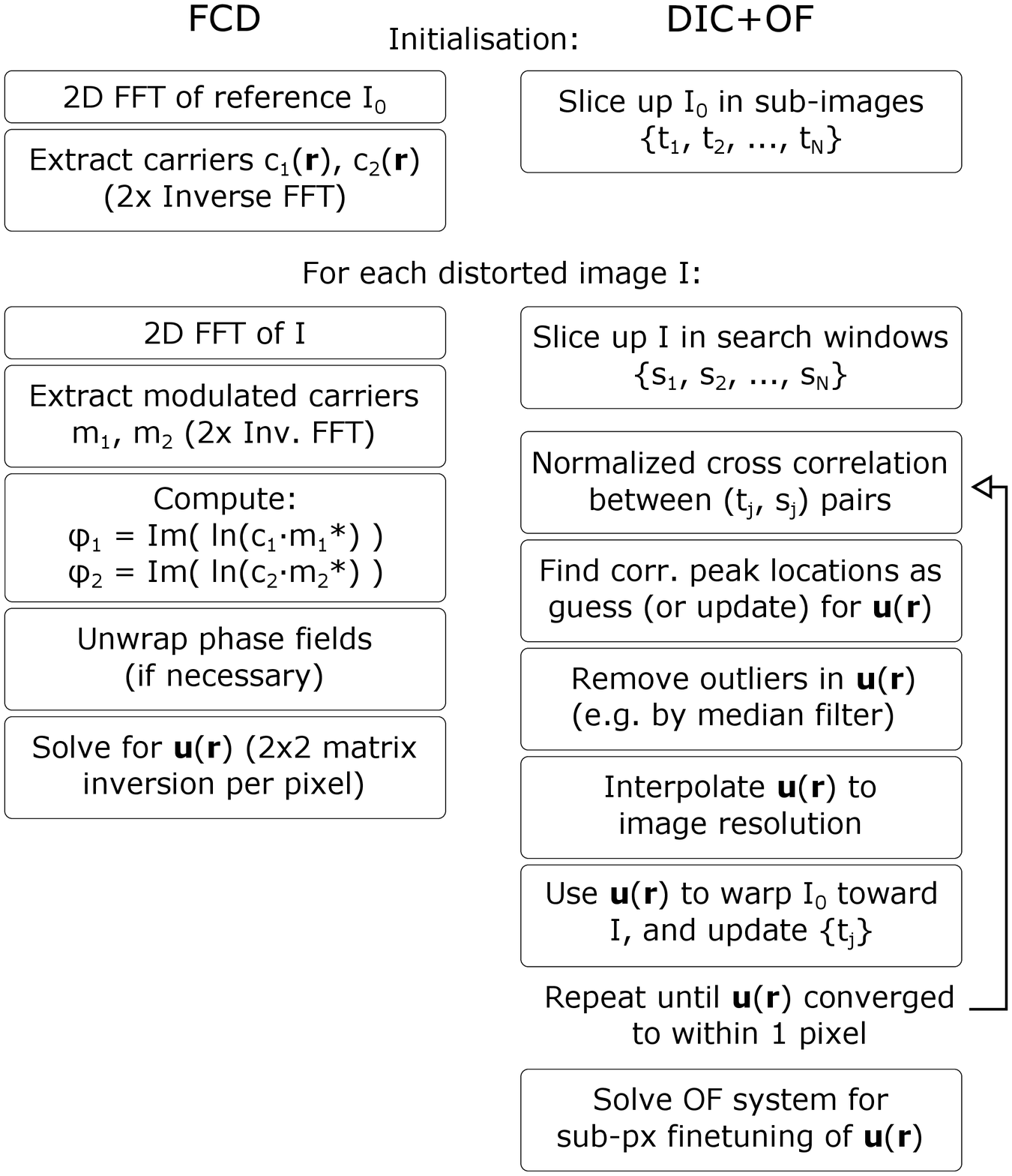}
\caption{Schematic outline of the computational steps involved in FCD (on the left) and DIC+OF (on the right).}
\label{fig:program}
\end{figure}

The distortions in the checkerboard images were analyzed using the FCD method described in section \sr{method} and the random dot images were processed with a hybrid DIC+OF scheme, akin to that proposed by Yang and Johnson \cite{Yang2017}. An outline of the computational steps involved in both schemes is given in Fig. \fr{program}. 

The hybrid DIC+OF scheme was specially implemented by the author for this comparison, as there seems to be no DIC/OF software (openly) available that is optimized for the optical distortion problem. Most current open DIC software is written with particle image velocimetry (PIV) or elastic strain measurement applications in mind and seem to employ iterative warping strategies that perform poorly on the optical distortion problem  (and hence would result in an unfair/biased comparison between DIC and FCD techniques for SSI). For reference, in Appendix \sr{pivlab}, PIVlab \cite{Thielicke2014} (a Matlab PIV code that is often employed in SSI applications) is tried on the soap bubble footage. 

The sub-image (or interrogation window) size used in the DIC analysis was $10\times10$ pixels, with a grid spacing of 6 pixels (corresponding to a window overlap of 20\%). The corresponding search areas, centered on the interrogation windows, had dimensions of $22\times22$ pixels. Cross correlations were performed using the fast normalized cross correlation algorithm proposed by Padfield \cite{Padfield2010}. Correlation peaks were located with sub-pixel accuracy using a basic $2\times3$ Gaussian peak fit \cite{Nobach2005}. After each iteration, the obtained vector field was interpolated to the size of the image and used to warp the reference image according to equation \er{warp}. This whole procedure was repeated with the updated reference until the displacement field converged to within 1 pixel. Usually two such iterations were sufficient. In the final step, for sub-pixel fine tuning, a Horn-Schunck OF system \cite{Horn1981} was solved with the regularization constant set to $\alpha = 0.03$.

Besides the displacement field, which is proportional to the gradient of the surface height $\nabla h(\vc r)$, one is often also interested in the surface profile $h(\vc r)$ itself.  This integral quantity can be obtained  either in real space, by a matrix inversion of a discretized gradient operator \cite{Snyder2002,Moisy2009,Harker2011,Cheng2012}, or in k-space, where the inversion consists of simple algebraic operations \cite{Frankot1988,Zhang1996,Bon2012,Huhn2016}. 

Because of the low computational cost associated with discrete Fourier transforms (DFTs), the Fourier integration method is generally significantly faster, and favorable when real-time visualization is desired (or large amounts of data need to be processed). However, care must here be taken to take into account the special boundary conditions (periodic, Dirichlet, or Neumann) implicitly assumed in DFT methods \cite{Bon2012,Huhn2016}. When displacements near the boundary of the image are significant, or in case some regions in the image are masked, the slower but more flexible real space inversion is usually preferred for accurate results \cite{Queau2017}. In the present case of localized disturbances generated by popping soap bubbles, both methods gave practically the same result.

Both the FCD and DIC+OF schemes were implemented as Matlab scripts (made available on GitHub \cite{github}). All steps were fully vectorized and only GPU-enabled functions of Matlab were used, so that the performance of both approaches could be tested on both CPU and GPU without modifications. All timing tests were performed with Matlab version 2016b running on a desktop PC with an Intel i7-6700 3.4 GHz processor and an NVIDIA Quadro M2000 video card with 768 CUDA cores.

\subsubsection{Measurement results}

Figure \fr{comp}(b) shows the displacement fields obtained with the FCD method (top half) and DIC+OF scheme (bottom half) (only the $x$ component, $u$, is shown here). At first sight, the results appear quite similar. In both cases, one can clearly observe how the initial pulled up rim ($t = 0$) splits into an outward and inward propagating wave ($t = 20\,$ms), of which the latter finally focuses in the center ($t=74$ms). However, upon more careful inspection one may notice that the FCD analysis somewhat better reproduces the small capillary waves propagating radially outward. This is especially clear at $t = 74\,$ms, where the displacement gradients are well within the range of both methods (condition \er{critus}). 

To probe the measurement accuracy in a more quantitative manner, the displacement fields were projected along radial rays originating from the center of the bubble. For each radial distance $r$, the mean $u_r(r)$ and standard deviation $\sigma(r)$ of the radial displacement were then determined, giving a measure of the deviation from the expected axial symmetry. The result of this analysis is plotted in Fig. \fr{comp}(c), where the thickness of the lines correspond to $4\sigma(r)$. Typical standard deviations obtained in this way are $\sigma_{\text{FCD}} = 0.02$ for the FCD method and $\sigma_{\text{DIC}} = 0.06$ for the DIC+OF scheme. Interestingly, the relatively high standard deviation seen around $r = 13\,$mm in both cases is due to a ring of smaller bubbles generated during the initial soap film collapse (see Fig \fr{comp}(a)) and is not related to measurement noise. Overall, the signal to noise ratio obtained with the FCD method was found to be roughly three times higher than that obtained with the DIC+OF scheme in this SSI setting. 

The surface profiles $h(\vc r)$ inferred from the measured FCD and the DIC displacement fields (as discussed in section \sr{impl}) are shown in Fig. \fr{comp}(d). The corresponding cross-sectional views (through the center of the bubble) are shown in Fig. \fr{comp}(e). To fix the integration constant, and to correct for small overall displacements (e.g., due to camera vibrations), a plane was subtracted from the obtained height fields such that the undisturbed border was leveled at $h = 0$. Because the integration involves solving an overdetermined system of equations in a least square sense, its effect is to smooth out any fluctuations inconsistent with the fact that $\vc u(\vc r)$ is a gradient. However, although the noise levels appear to be equally small in the height fields recovered from the FCD and DIC analysis, the smallest capillary waves are only recovered in the FCD analysis (see inset Fig. \fr{comp}(e)).

\begin{figure}
\centering
\includegraphics[width=\textwidth]{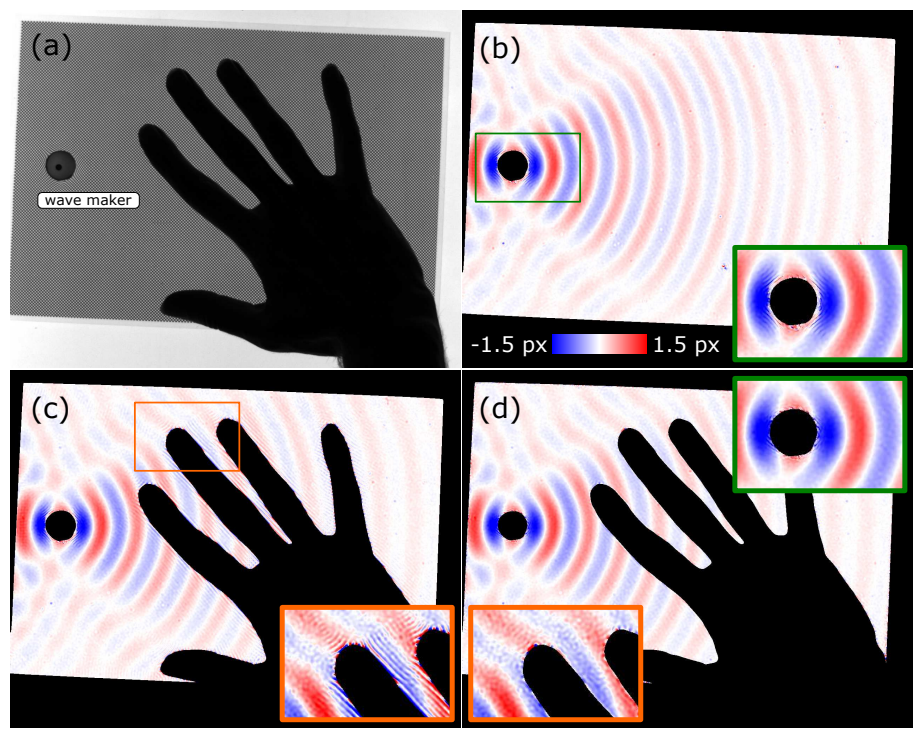}
\caption{Effect of occlusions on the FCD analysis. (a) Top view of the experiment. A small oscillating floater (wave maker) generates waves at the surface of a pool of water. A checkerboard pattern is placed below the pool for performing FCD analysis. The experimenter's hand is blocking part of the view. (b) Result of FCD analysis without the hand occlusion (the horizontal displacement component is shown). Since the mask (black regions) did not undergo any treatment, small fringe artifacts are visible near the wave maker (see inset). (c) Similar artifacts appear around the edge of the hand once it is brought into view (see inset for a magnification with enhanced contrast). (d) The masking artifacts are removed by low pass filtering the wave maker's mask and filling-in the hand's mask with the undistorted reference pattern before performing the FCD analysis.}
\label{fig:masking}
\end{figure}

In the snapshot at $t=0$ in Fig. \fr{comp}(e), it seems that the FCD and DIC analysis recover completely different water heights below the bubble before it pops. In fact, compared to the theoretical prediction \cite{Pujado1972,Ross1992} shown in the same plot, both methods give a height which is much too high (the surface tension of the soap solution was estimated to be $\sigma = 25\,$mN/m \cite{Roman2001}). This discrepancy occurs because near the edge of the bubble, the water surface becomes vertical (and even multi-valued at some points). The measured displacement fields can clearly not be trusted here. Effectively, the domains inside and outside the bubble are completely disconnected and will therefore have different (more or less arbitrary) integration constants. Notice, however, that just after the bubble has popped, and before the waves reached the center (at $t=20\,$ms), the predicted surface depression is recovered by both methods.

Besides the lower noise level, the main advantage of FCD over DIC is probably in its algorithmic simplicity, and related to that, its speed. On the PC used for the analysis, a sequence of images of $512\times512$ pixels could be processed at a frame rate of 50 FPS on the CPU and 190 FPS on the GPU, easily fast enough for providing a real-time quantitative preview. This has to be contrasted with 0.5 FPS and 1 FPS, respectively, for the DIC+OF scheme.

\subsection{Occlusions and masking}

In Schlieren imaging, it is often unavoidable that parts of the scene are occluded, either by static objects, which block the view to both the reference and the distorted pattern, or by moving objects, which only appear in the distorted image. As in DIC \cite{Padfield2010}, some care has to be taken within the FCD framework to prevent these masked regions from spoiling the displacement field reconstruction in the remaining part of the image (especially near the borders of the occlusions).

In Fig. \fr{masking}(a), a snapshot from a water wave experiment is shown in which the view to the background pattern is partially blocked by a small circular wave maker (which is present in both reference and distorted images) and by the hand of the experimenter (which was brought into view after the recording was started). In addition, the regions not covered by the checkerboard pattern are effectively occluded.

Figure \fr{masking}(b) and (c) show two displacement fields, one with hand and one without hand, which were obtained without any special treatment of the occluded areas during the FCD analysis (the black masks were applied afterwards for clarity of presentation). One can observe that away from the mask borders a faithful reconstruction is obtained in both cases, while close to the wave maker and in between the finger tips (see insets) some fringe-like disturbances are visible. This shows that although part of the FCD analysis has a global character (namely the filtering in the Fourier domain), the artifacts caused by a local mask remain local in the final result.

The appearance of fringe-like artifacts around occlusions can be understood as follows.  According to the convolution theorem of Fourier analysis, the effect of multiplying the pattern by a mask (also called a window function) is to convolve the Fourier spectrum of the pattern by the Fourier transform of the mask. Since the pattern's spectrum is sharply peaked around the four carrier wave vectors, this means that four copies of the mask's spectrum, centered on the carrier peaks, are superimposed on the Schlieren signal. In principle this is not a problem, as in the final step of the analysis (after filtering) the reverse operation occurs and the mask simply multiplies the extracted complex signal in real space. However, if the mask's spectrum contains spatial frequencies that are higher than $\sim k_c/\sqrt{2}$, then these frequencies are not properly recovered and one observes an aliasing effect. Since the amplitude of the mask spectrum will generally decay with increasing $k$, the strongest artifacts will have a wavelength of the order of $2\pi/k_c$ (i.e. the wavelength of the background pattern) as is clearly visible in Fig. \fr{masking}(c).

From the above analysis, a straightforward solution presents itself to prevent masking artifacts: instead of directly applying a binary mask to the reference and distorted images, one should apply a low-pass filtered mask, from which spatial frequencies $> k_c/\sqrt{2}$ are removed (i.e., apply an anti-aliasing filter). This can be efficiently implemented as a dilation the original binary mask, followed by a convolution with a 2D Gaussian kernel of width $\sim 2\pi/k_c$. Another effective method, applicable to moving occlusions, is to simply fill-in the masked regions with the undistorted pattern from the reference image. 

In Fig. \fr{masking}(d), the anti-aliasing approach was applied to the mask of the wave maker, while the filling-in approach was applied to the mask of the hand. As can be seen in the insets in Fig. \fr{masking}(d), most masking artifacts are removed after these two operations.

\section{Conclusion and outlook}

The spatial Fourier demodulation technique was extended to arbitrary 2D periodic patterns. This simple framework, coined FCD for Fast Checkerboard Demodulation, allows for quantitative extraction of the full vector displacement field in synthetic Schlieren imaging applications when a regular, checkered background is used. It was found that the method can resolve large virtual strains, on par with conventional image correlation and optical flow techniques, with a favorable dynamic range (about three times larger than the tested DIC+OF scheme). Furthermore, information on whether a good reconstruction can be expected can be directly obtained by inspecting the signal in the Fourier domain, which can be helpful for data validation. Due to the high computational efficiency of the fast Fourier transform used in the FCD method, real-time processing (typically 50 FPS at an image resolution of $512 \times 512$ pixels) can be achieved on a standard desktop PC. This allows for live data previewing while setting up an experiment (akin to the direct visual feedback in classical Schlieren systems, but with the advantage of being quantitative) and facilitates high throughput data processing afterward.

In the present work, background patterns with a square lattice were used with the FCD analysis. However, the procedure described in section \sr{method} is general and can be used with any type of periodic background. One could, for example, imagine that there are experimental situations in which more bandwidth is required in one direction compared to another. In this case, it could be advantageous to use a rectangular grid, with a small lattice spacing in one direction and a larger one in the other. Another option would be to use a pattern with multiple frequency components. The lower frequency component could then, for example, serve as a guide to unwrap the signal obtained from the denser grid \cite{Fu2013,Hatanaka2015,Yun2017}. 

Furthermore, the reference images in this study were obtained simply by capturing an image when the water surface was still. However, in principle, the regular reference pattern could also be generated digitally. An interesting application of this is that, by performing FCD analysis between this generated, ideal reference and an image of the (unperturbed) physical background pattern, one can obtain detailed information on the alignment of the background pattern and possible lens distortions. This could for example be useful for \textit{in situ} calibration of the optical setup. In addition, in situations where no undistorted reference image can be readily obtained (e.g., in monitoring industrial heat flows) this feature can be useful.

To conclude, when one has the freedom to choose the type of background pattern to place behind the Schlieren object, FCD in combination with a regular 2D periodic pattern offers a flexible, accurate and fast quantitative SSI alternative to the conventional DIC and OF techniques.

\begin{acknowledgements}
I am indebted to Antonin Eddi and Emmanuel Fort for their enthusiastic support during the development of the FCD method in their groups. I would also like to thank Guillaume Du Moulinet d'Hardemare and Lucie Domino for testing the method in their experiments and giving the necessary practical feedback.
\end{acknowledgements}


\section*{Appendix: Comparison to PIVlab \label{sec:pivlab}}


\begin{figure}
\centering
\includegraphics[width=\textwidth]{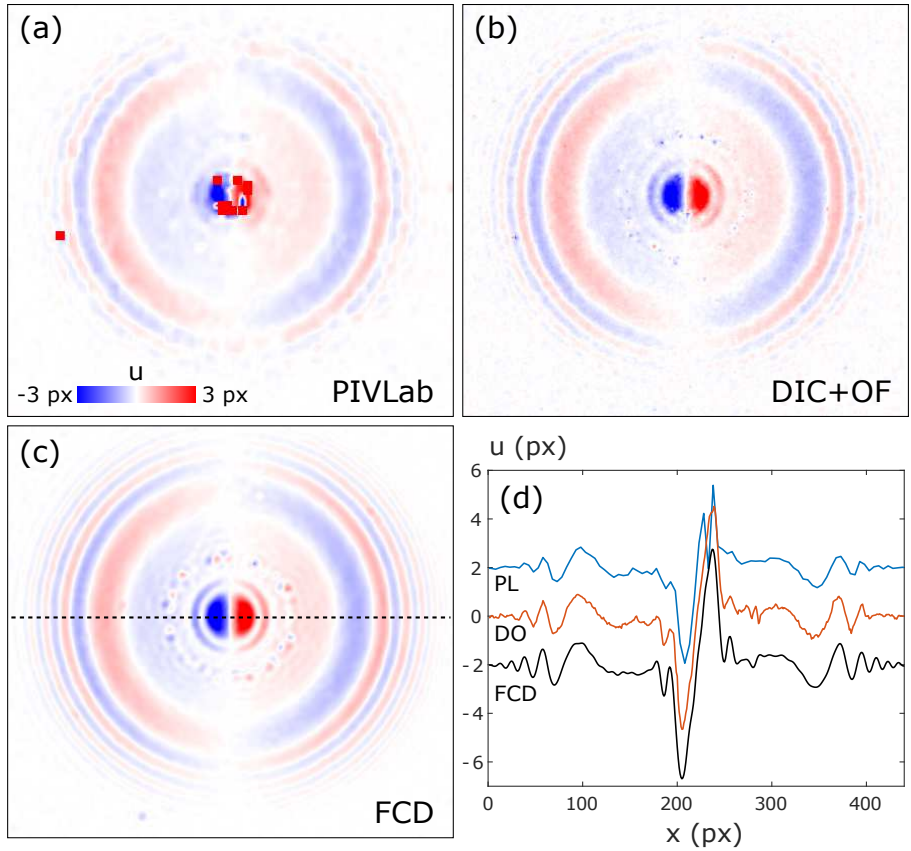}
\caption{Comparison between displacement fields obtained by (a) PIVlab, (b) DIC+OF and (c) FCD operating on the soap bubble data. Only the horizontal displacement component is shown here. In (d), horizontal traces through the center of the bubble, as indicated by the dashed line in (c), are plotted. To facilitate comparison, the curves in (d) were slightly offset from each other}
\label{fig:pivlab}
\end{figure}

 To perform DIC analysis in synthetic Schlieren applications, often out-of-the-box PIV software, developed for the tracking seeding particles in a flow, is employed. However, the imaging of optical distortions presents somewhat different challenges than the tracking of particles in a flow. It can therefore be expected that the performance of PIV optimized software is not necessarily optimal for SSI. 

In PIV systems, for example, particle seeding conditions are often far from optimal, favoring robustness of the correlation algorithm over accuracy. In a typical SSI setup, on the other hand, one can simply create the perfect particle pattern, so that any loss in accuracy becomes directly apparent. Furthermore, particle displacements caused by a flow are often closer to the basic translation-only assumption underlying DIC, so that window deformations can be incorporated as a second-order effect (also the particles themselves do not deform in PIV). In contrast, in Schlieren imaging, the virtual strains (at the scale of the texture) are often very large.

To illustrate the above points, and to justify the use of a custom-made DIC+OF scheme in the comparison in the main text, a short comparison to the openly available PIVlab software (v. 1.43) \cite{Thielicke2014} is included here. In Fig.~\fr{pivlab}, the horizontal distortions obtained by (a) PIVlab, (b) DIC+OF and (c) FCD for the popping bubble experiment at $t = 74\,$ms are shown. Fig. \fr{pivlab}(d) shows horizontal traces through the center of the bubble for each case. PIVlab and DIC+OF were fed images from the same experiment with a random dot background, while the FCD data was obtained in a separate experiment with a checkerboard background. In PIVlab, the three-stage multi-pass FFT cross correlation option was used, which applies a window deformation between each stage. The correlation window sizes for each stage were set to of $16\times16$, $10\times10$ and $10\times10$ pixels, respectively. PIVlab performs cross-correlation between equally sized window pairs \cite{Thielicke2014}, as opposed to the DIC+OF scheme, in which each sub-image in the reference is correlated with a larger search area in the distorted image.

Comparing Fig. \fr{pivlab}(a) and (b) it is immediately clear that the DIC+OF method recovers the image displacements with significantly higher fidelity than PIVlab. Even after two window deformation cycles, PIVlab fails to resolve the strong optical distortions at the center of the image (showing NaN's in that region). Furthermore, as is seen in Fig. \fr{pivlab}(d) PIVlab overall reports somewhat smaller distortions than the DIC+OF and FCD schemes, and does not recover the smallest capillary waves. This is probably due to PIVlab's bias to zero displacement, inherent to the correlation technique it employs (this underestimation effect can be reproduced by artificially shifting the reference by 1 pixel and then performing the analysis between the unshifted and shifted version, resulting in a reported mean displacement of 0.95$\pm 0.06$ px by PIVlab, while DIC+OF correctly returns 1.00$\pm 0.04$ px). Although PIVlab returned its result roughly five times faster than DIC+OF (although at a five times lower resolution), in terms of both speed and accuracy, the two correlation methods are still greatly outperformed by the FCD method (which was roughly 16x faster than PIVlab).

\bibliographystyle{spphys-fulltitle}       
\bibliography{fcdrefs}   

\end{document}